# Accretion powered AGN feedback in the cores of galaxy clusters

**M K Patil**
School of Physical Sciences
S.R.T.M.University, Nanded – 431 606, Maharashtra

**Abstract:**

Detection of the copious amount of X-ray emission from the dilute hot plasma in galaxy clusters suggests that a substantial fraction of the central intracluster medium (ICM) is cooling radiatively on a time scale much faster than the Hubble time. Theoretical models predict the cooling rate as high as about few hundred to few thousand solar mass per year, which would be then made available for the formation of new stars in the core of these clusters. However, systematic studies of the cores of such clusters failed to detect the expected reservoirs of cooled gas. Thus, the gas in the cores of galaxy clusters is losing substantial amount of energy in the form of X-rays but is not cooling. This in turn point towards the famous cooling flow paradox and hence demands some intermittent heating to balance the cooling over such a long period. Several sources have been suggested to counteract on the cooling of the ICM, however, the AGN feedback appeared to be the most promising and enough energetic source to resist cooling of the ICM in the cores of such clusters. In this presentation I will provide a brief overview on the feedback processes that are involved in the cores of the galaxy clusters with an emphasis on the AGN feedback and its observable signatures.

## 1. Introduction

Galaxies are the building blocks of the universe and are comprised of a variety of objects that include, planetary systems, stars, associations of stars, interstellar dust, gas, and importantly a supermassive black hole at the center. They come in a variety of forms, shapes and sizes such that, the smallest galaxies may contain a few hundred thousand stars, while the largest may host billions of stars. They also come in different flavors like, ellipticals, spirals, lenticulars, and irregulars. The most massive galaxies are the giant ellipticals, million times massive than the smallest with masses as high as several trillion solar mass. These galaxies are normally found in the cores of the groups and clusters of galaxies. As far as distribution of galaxies in the universe is concerned, they are not randomly distributed in the space as isolated mass aggregations, but normally exist in groups of a few to few dozens of galaxies or large clusters of up to several thousand galaxies. Galaxy clusters have typical masses in the range between $10^{13}$ - $10^{15}$ $M_{sun}$ and span over about few Mpc. The largest fraction of the mass (about 75%) of a galaxy cluster is in the form of dark matter. Only 25% of the mass is in the form of ordinary matter, out of which a small fraction (about 5%) has condensed in the form of stars, cold gas and dust (Spergel et al. 2003). Thus, clusters of galaxies are the most massive, relaxed, gravitationally bound systems in the universe and act as most prominent density peaks that hold clues regarding formation and evolution of the universe.

In contrast to the optical observations of the galaxy clusters, where they appear as localized over densities of hundreds to thousands of galaxies tightly concentrated on the plane of the sky, in X-ray bands galaxy clusters appear as a single, X-ray bright, extended source with luminosity ranging between $10^{43}$ - $10^{45}$ erg s$^{-1}$. This means, the huge space between the constituent galaxies in a cluster is not empty but is filled with a copious amount of hot, diffuse gas heated to temperature $10^7$ - $10^8$ K and is known as intra-cluster medium (ICM). The ICM was believed to be heated to such a high temperature due to its adiabatic compression because of the gravitational in fall. Recent deep, high resolution observations of the galaxy clusters using instruments like Chandra X-ray observatory have detected sharply peaked surface brightness profiles in several of them, indicating that the gas density rises steeply in the cores of the galaxy clusters (Fabian 1994, Gitti et al. 2012, Pandge et al. 2012, 2013, Vagshette et al. 2016). Higher values of X-ray flux densities from the core of the galaxy clusters indicate that the hot gas is losing most its energy in the form of X-rays. As a consequence, to maintain the pressure balance in the central region, a sub-sonic inflow of gas would start and hence will lead to the so called cooling flow of the ICM (Fabian 1994). Theoretical studies demonstrate that such a balance requires deposition of about 100 - 1000 Msun of the ICM on to the core of such cluster (Fabian & Nulsen 1977, Mathews & Bregman1978). This means cores of such clusters must be having reservoirs of about 100 - 1000 Msun of cool gas, which will then be utilized for the new star formation. However, recent high-resolution X-ray observations from Chandra and XMM-Newton telescopes failed to detect expected amount of the condensed gas in the cores of the cooling flow clusters. Even the high-resolution spectral capabilities of XMM-Newton telescope also failed to detect the predicted spectral features of the cooling gas (Peterson et al. 2001, 2006). Further, observations at other wavelengths from instruments like HST could not find evidences of the large reservoirs of cold gas at the centers of cooling flow clusters. The systematic discrepancy between the cooling rate of X-ray emitting gas and the absence of equivalent amount of condensed gas in the cores lead to the famous "*cooling flow problem*" of the cool core clusters (Fabian 2012).

Systematic studies of cooling flow galaxy clusters employing high-resolution X-ray imaging and spectroscopic observations acquired using the Chandra and XMM-Newton space telescopes have figured out that the in falling gas is not cooling directly from the plasma phase to the molecular phase, but is instead reheated before condensing (Peterson et al. 2006). The current paradigm is that the AGN residing at the core of cluster dominant galaxy is responsible for such a re-heating of the cooling gas, implying a more complicated feedback process between the cooling gas and the central galaxy. It also plays an important role in shaping the morphologies of hot gas halos surrounding individual galaxies in groups and clusters and result in to the formation of cavities and bubbles (Birzan et al. 2004, Dunn & Fabian 2006, McNamara & Nulsen 2007, Dong et al. 2010, Pandge et al. 2012, 2013, Sonkamble et al. 2015, Vagshette et al. 2016). X-ray cavities are nothing but the depressions in the surface brightness distribution of X-ray emission and are believed to be filled with low density relativistic plasma, and provide the direct evidence of the interaction between the central AGN and the surrounding gas (Fabian 2012). Now it is widely accepted that mass accretion onto the central massive black hole or spin of the SMBH could be the source of energy for the formation of these cavities. Therefore, detailed investigations of X-ray cavities in the cooling flow systems are proxy to understand the complicated AGN feedback and also to solve the cooling flow problem.

## 2. Evidences of AGN feedback

The morphological analysis of the X-ray emission distribution in these clusters employing data from Chandra X-ray telescope has provided us with the amazing details regarding interplay between the radio jets emanating from the central AGN with the surrounding plasma by carving X-ray cavities, shocks, cold fronts, surface brightness edges, ripples, turbulence, gas sloshing etc. It has been observed that the surrounding ICM regulates expansion of the radio plasma thereby transferring a large amount of energy to the ICM, which in turn suppresses cooling of the ICM. The direct apparent evidences due to such an interaction are provided in the form of cavities, filaments and edges in the surface brightness distribution and have been confirmed though a variety of image processing techniques like, unsharp masked, 2D β-model subtracted residual images, subtracting smooth models generated using the GALFIT or ellipse fitting analogies, etc. This analysis has enabled us to detect X-ray cavities in nearly all the cool core galaxy clusters. Fig. 1 (left) delineates such a detection in the environment of 3c344, where a pair of X-ray cavities can be clearly evidenced. These cavities are found to be surrounded by rims of compressed cool gas. It is believed that the radio jets emanating from the radio sources associated with these clusters can evacuate cavities or bubbles in the X-ray surface brightness distribution. Like the X-ray cavities, radio lobes also appear in pairs on either sides of the central source. Fig. 1 (right) reveals such a case for 3c444 source, where X-ray cavities (shown in blue) appear to be filled in by the radio emission (shown in green) originating from the central engine (Vagshette et al. 2017).

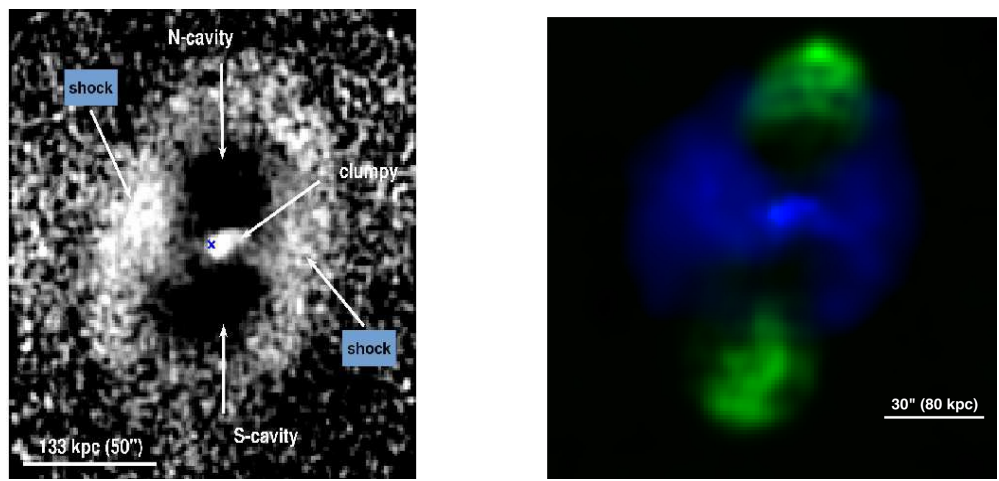

Fig. *(left)* 0.5-3 keV Chandra X-ray unsharp masked image of 3c444 delimiting a pair of large X-ray cavities surrounded by cool rims. *(right)* Spatial associations of the X-ray cavities (blue) and the radio jets (green), confirming that cavities are carved by the radio jets.

In addition to the X-ray cavities this presentation also reports detection of several other signatures of such interactions apparent in the form of sharp discontinuities due to shocks, cold fronts, ripples, edges, gas sloshing, etc. The observed shocks and cold fronts have resulted in to the sudden jump in the measured values of ICM temperature.

2D spatially resolved spectral analysis of the X-ray emission from these systems enabled us to investigate the radial nature of the thermodynamical profiles (kT, ne, p, S) and has also enabled us to investigate sudden jumps in the profiles of several of the thermodynamical parameters, implying that the gas in those regions is either compressed or is suddenly shocked. We find that such jumps in the temperature, electron density and pressure are due to the presence of cold fronts and shocks in the wake of the hot gas environment. These profiles are important tool to probe the galaxy clusters in the sense that, the temperature profile represents the depth of the cluster potential, while the capacity of the well to compress the gas can be assessed though the hot gas density profile.

## 3. Entropy as the probe to gravitational potential

The entropy is another very important entity that holds information regarding the thermal history of the ICM as the outcome of both gravitational as well as non-gravitational processes occurring at the cores of such clusters (Voigt et al., 2004), which changes only when there is either a gain or loss of energy from the system. Thus, the entropy profiles for cluster provide an important tool for investigating the effects of AGN heating and radiative cooling. The radial entropy profiles derived for all of the systems studied here revealed a "floor" or "ramp" in the central region, which then rises systematically in outer direction, following the classical power-law. However, a significant deviation from this power-law nature in the central region is consistent with the observations by several researchers and implies that some intermittent heating is operative in the core of the clusters studied here.

Once it was established that some intermittent heating is operative in the core of these clusters, systematic attempts were made to investigate the possible source for such a heating. Several sources were suggested in this regard, however, all these failed to prove their capability to meet the requirement at such a large scale over a very long time. Therefore, finally it was felt that the possible source for heating of the ICM in the core of these clusters cold be none other than the AGN feedback.

## 4. X-ray cavities as calorimeters

Once it is confirmed that AGN feedback can play an important role in balancing cooling and heating of the ICM, we made attempt to quantify the amount of mechanical power that has been delivered by the AGN to the surrounding ICM. This was achieved by employing buoyantly rising X-ray cavities as the calorimeter. Here, we believe that the kinetic energy of jets emanated from the central black is greater than the amount of energy required to inflate the X-ray cavities (Birzan et al. 2004). This study has revealed that, on an average, X-ray cavities in the cool core clusters can deliver mechanical energy in the range between $2.30 \times 10^{44}$ to $6.13 \times 10^{45}$ erg s$^{-1}$ and is produced by the AGN through the process of the accretion of about $10^7$ to $10^9$ Msun of matter on to the central supermassive black holes (SMBH) over the period of about few Gyr. The balance between the injected mechanical energy ($P_{cav}$) versus that has been lost by the ICM in the form of radiative loss ($L_{cool}$) is shown in Fig. 2, which reveals that the central AGN is capable enough to quench the cooling of the ICM at least in the cores of these clusters.

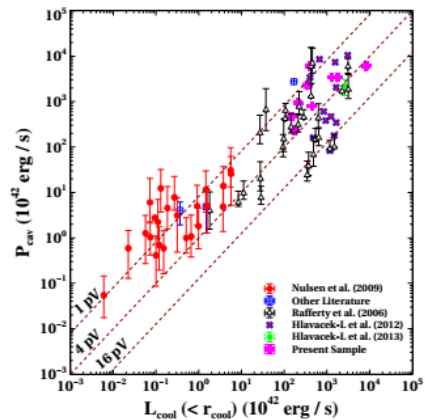

Fig.2 Balance between the mechanical power injected by AGN versus radiative loss of the ICM.

Thus, AGN feedback can regulate the cooling of the ICM in the cores of the galaxy clusters and can suppress the star formation. At the same time it leaves the impacts of such an interaction without erasing them. There are still some evidences of multiphase gas association in the cores of few clusters. The cold gas with T ≤ $10^4$ K is generally seen in the form of spatially extended clouds or filaments in few of the systems from our study. However, due to lack of high quality optical narrow band imaging as well as spectroscopic data on other systems did not allow us to check similar possibilities in rest of the systems.

**5. Conclusions:**

This presentation discusses multifrequency observations of the central dominant galaxies hosted by the cool core clusters with an emphasis to investigate depressions or cavities in the X-ray surface brightness distribution of the ICM. It has been observed that the interplay between the radio jets originating from the central dominant galaxies and the surrounding ICM regulates expansion of the radio plasma and transfer large amount of energy to the ICM and thereby suppress cooling of the ICM. Several other signatures due to such interactions were also evidenced in the form of sharp discontinuities due to shocks, cold fronts, ripples, edges, gas sloshing, etc. It has also been demonstrated that the accretion powered central AGNs are the only plausible sources to regulate cooling of the ICM and hence its evolution.